\documentclass[prl,twocolumn,a4paper,aps,showpacs,superscriptaddress]{revtex4}

\usepackage{graphicx}% Include figure files
\usepackage{dcolumn}% Align table columns on decimal point
\usepackage{bm}% bold math
\usepackage{amssymb}

\begin{document}
\title{Evidence of correlation in spin excitations
of few-electron quantum dots}
\author{C\'{e}sar Pascual Garc\'{i}a}
\affiliation{NEST-INFM and Scuola Normale Superiore, Piazza dei
Cavalieri 7, I-56126 Pisa, Italy}
\author{Vittorio Pellegrini}
\affiliation{NEST-INFM and Scuola Normale Superiore, Piazza dei
Cavalieri 7, I-56126 Pisa, Italy}
\author{Aron Pinczuk}
\affiliation{Dept of Physics, Dept of Appl.~Phys.~and Appl.~Math.,
Columbia University, New York, New York} \affiliation{Bell Labs,
Lucent Technologies, Murray Hill, New Jersey}
\author{Massimo Rontani}
\author{Guido Goldoni}
\author{Elisa Molinari}
\affiliation{S3-INFM and Dipartimento di Fisica, Universit\`a degli
Studi di Modena and Reggio Emilia, 41100 Modena, Italy}
\author{Brian S. Dennis}
\author{Loren N. Pfeiffer}
\author{Ken W. West}
\affiliation{Bell Labs, Lucent Technologies, Murray Hill, New
Jersey}
\date{\today}
\begin{abstract}
We report inelastic light scattering measurements of spin and
charge excitations in nanofabricated AlGaAs/GaAs quantum dots with
few electrons. A narrow spin excitation peak is observed and
assigned to the intershell triplet-to-singlet monopole mode of
dots with four electrons. Configuration-interaction theory
provides precise quantitative interpretations that uncover large
correlation effects that are comparable to exchange Coulomb
interactions.
\end{abstract}
 \maketitle
\par
Electrons confined to semiconductor quantum dots (QDs) have novel
ground and excited states that manifest Coulomb interactions at
the nanoscale \cite{Reimann}.  States of very few electrons
are prime candidates for spintronic applications and for the
implementation of quantum bits
in nanoscale devices \cite{loss98}. Great attention
is therefore devoted to the study of spin physics in the regime of
few-electron occupation and to experimental methods
capable of reading the state of spin in the QD \cite{kouw04}.
\par
The interpretation of experiments on few-electron QDs often
requires descriptions beyond mean-field, such as Hartree-Fock (HF)
\cite{Reimann,haw04}. In addition to their relevance for quantum
information encoding, these correlated states 
have significant interest for the investigation of fundamental
effects \cite{egger,oliver}. Transport in few-electron QDs coupled to 
leads and excitonic optical recombination measurements have
explored exchange energies and spin relaxation times. These
remarkable experiments offer evidence for roles played by
interactions that emerge as the number of electrons in the QD is
changed \cite{Taruch,exchange,kouwe01,fuji02,kogan03,kouw04,
petta04,zhumbul04,haw04,bayer98,fin04}.
\par
In this Letter we report resonant inelastic light scattering experiments in low electron 
density GaAs/AlGaAs QDs that probe low-lying {\it neutral} excitations. 
These are inter-shell monopole excitations (with change in Fock-Darwin (FD) shell 
and without change in angular momentum, as required by light scattering selection rules \cite{lockwood,schuller}). We detected two broad inter-shell modes that we interpret 
as excitations of electrons from the two populated lowest shells.
Each of these two modes is split by exchange
and depolarization effects into a $\Delta S =1$ (spin) and a $\Delta S=0$ (charge) excitation \cite{lip,delg}, where $\Delta S$ represents the change of the total spin of the QD associated to the electronic mode.
\par
A prominent feature of the spectra is an additional mode peculiar to the regime of few-electron occupation that emerges at low temperature and low excitation power. It occurs as a very narrow peak with light scattering polarization selection rules for spin excitations and is interpreted as a $\Delta S=-1$ intershell spin mode characteristic of a $S=1$ triplet ground state with four electrons. We argue that the observed splitting between the $\Delta S=1$ and $\Delta S=-1$ spin modes represents a direct manifestation of the role of interactions in the excitation spectra of few electron QDs. 
\par
Numerical evaluations within a configuration-interaction (CI) theory \cite{ront04} support the interpretation that links the new spin mode to the triplet-to-singlet (TS) inter-shell excitation of a QD with four electrons and offer quantitative insights on the role of interactions in this regime. The CI evaluations reproduce the experimental light scattering spectra with a great precision that is not achieved by HF theory. Comparisons of mean field and CI calculations uncover large exchange and correlation terms of electron interactions that in the case of the four-electron triplet state are found to be comparable to quantum confinement energies.
\par
\begin{figure}
\includegraphics[width=7cm]{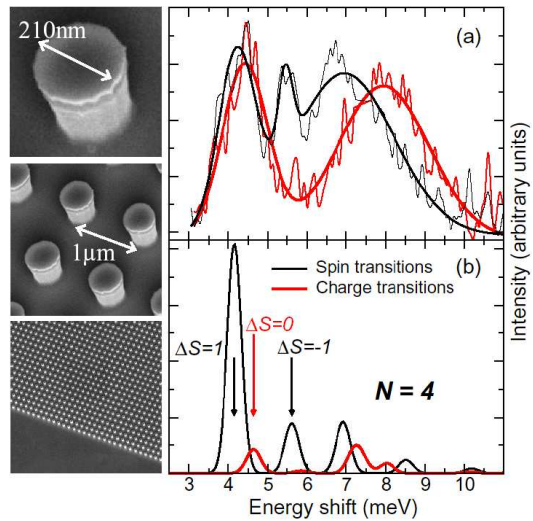}
\caption{\label{fig1} Side panels: Scanning electron microscope
images of the studied QDs. (a) Experimental low-temperature ($T=1.8$ K)
polarized (parallel incident and scattered photon polarizations,
in red) and depolarized (perpendicular incident and scattered
light polarizations, in black) resonant inelastic light scattering
spectra after conventional subtraction of the background due to laser light.
Incident laser energy is 1567 meV, intensity is I = 0.08 W/cm$^2$
and integration time is 30 minutes. Fits of
the data with three gaussians are shown. (b) Theoretical spectra
for electron number $N = 4$. Gaussian distributions with a 
phenomenological standard deviation $\sigma = 0.18$ meV were used.}
\end{figure}

Samples were fabricated from a 25 nm wide, one-side
modulation-doped Al$_{0.1}$Ga$_{0.9}$As/GaAs quantum well with
measured low-temperature electron density $n_e = 1.1\times 10^{11}$
cm$^{-2}$ and mobility of $2.7\times 10^6$ cm$^2$/Vs. QDs were
produced by inductive coupled plasma reactive ion etching. QD
arrays (with sizes $100\times 100$ $\mu$m containing $10^4$ single
QD replica) were defined by electron beam lithography with
different diameters. Deep etching (below the doping layer) was
then achieved. Here we focus on QDs having lateral
lithographically-defined diameters of 210 nm (shown in
Fig.~\ref{fig1} side panels) that we expect to be close to the
regime of full electron depletion \cite{kir02}. The experiments
were performed in a backscattering configuration ($q\le 2\times
10{^4}$ cm$^{-1}$ where $q$ is the wave-vector transferred into
the lateral dimension) with temperatures down to $T=1.9$ K. A
tunable ring-etalon Ti:sapphire laser was focussed on 100 $\mu$m-diameter
area and the scattered light was collected into a triple grating
spectrometer with CCD multichannel detection.
\par
A convenient description of single-particle QD levels is provided
by FD orbitals \cite{Reimann} with energies given by 
$\varepsilon_{nm}= \hbar \omega_0 (2n +
|m| + 1)$, where $n=0,1,\ldots$, $m=0,\pm 1, \ldots$ are
the radial and azimuthal quantum numbers, respectively, and
$\hbar\omega_0$ is the harmonic confinement energy. The FD shells
are defined by an integer value of $N_{\text{shell}} = 2n+|m|$
with well defined atomic-like parity. QD states can be classified in terms of
the $z$-component  total angular momentum $M$, total spin $S$, and
its $z$-component $S_z$. Selection rules in QDs dictate that
the monopole transitions with $\Delta M = 0$ ($\Delta N_{\text{shell}}
=2,4,\ldots$) are the inter-shell modes active in light scattering
experiments \cite{strenz,schuller,lockwood}. This
non-interacting picture of intershell transitions is shown in the
left part of Fig.~4, where the lowest energy dipole
($\Delta N_{\text{shell}}=1$)
and monopole ($\Delta N_{\text{shell}}=2$) modes are represented.
\par
Figure \ref{fig1}(a) shows representative low-temperature spectra
of inter-shell spin and charge excitations that are detected with
crossed and parallel polarization between incident and scattered
light, respectively \cite{vit01}. Pairs of peaks are seen at
energies close to $4$ meV and $7-8$ meV and interpreted 
as monopole excitations with $\Delta S=1$ (spin)
or $\Delta S=0$ (charge). In the non-interacting FD picture these 
two excitations are degenerate but they split in the presence of 
exchange and depolarization contributions. 
A characteristic feature of these doublets is
their significant linewidth that is attributed largely to
inhomogeneous broadening due to the distribution of electron
occupations of the dots as described below. 
\par
\par
The spectra in Fig.~1(a) reveal a much sharper excitation in the
spin channel at $5.5$ meV. To interpret the origin of this sharp
spin mode we note that an additional TS intershell spin mode with
$\Delta S=-1$ can occur if the ground state is a triplet with $S=1$
and the excited state is a $S=0$ singlet state (see the calculated
levels shown in Fig.~4). Such TS excitation is split from the
$\Delta S= +1$ mode seen at lower energy by the difference in
exchange and correlation contributions. On this basis we identify
the sharp peak at $5.5$ meV with the TS ($\Delta S=-1$) intershell
spin excitation. According to Hund's rules a triplet ground state
occurs only when two electrons are in a partially populated
shell as is the case of QDs with four electrons. 
This is confirmed by the calculations described below.
In this interpretation the narrow width is simply explained as due to
the absence of inhomogeneous broadening from the distribution of the
electron population of the QDs.
\par
We used the full CI approach \cite{ront04,Haw95,method} for the numerical
evaluation of the energy and intensity of low-lying spin and charge
excitations of the interacting system with $N$ electrons.  The
correlated wavefunctions of ground and excited states are written as
superpositions of SDs,
$\left|\Phi_{\{\alpha_i\}}\right>=\prod_{i=1}^N c^\dag_{\alpha_i}
\left|0\right>$, obtained by filling in the single-particle
spin-orbitals $\alpha$ with the $N$ electrons in \emph{all} possible
ways [$c^{\dagger}_{\alpha}$ creates an electron in level
$\alpha\equiv (n,m,\uparrow \text{or} \downarrow )$]. The resulting
Hamiltonian is first block diagonalized, fully exploiting symmetries
\cite{method}. Finally, the Hamiltonian is diagonalized via
Lanczos method in each $(M,S,S_z)$ sector, giving the low-energy
excited states. The resonant Raman transition matrix elements
$M_{\text{FI}}$ between the fully interacting ground and excited
states $\left|\text{I}\right>$ and $\left|\text{F}\right>$,
respectively, are obtained, after the CI calculation, from $
M_{\text{FI}} = \sum_{\alpha\beta} \gamma_{\alpha\beta} \left<
\text{F} | c^{\dagger}_{\alpha} c_{\beta} | \text{I} \right>$, where
$\gamma_{\alpha\beta}$ is the two-photon process matrix element
between $\alpha$ and $\beta$ spin-orbitals, as defined in
Ref.~\onlinecite{Steinebach} within second order perturbation theory
in the radiation field and containing resonant denominators.
$\gamma_{\alpha\beta}$ causes the enhancement of the spectrum
intensity when the laser energy resonates with the optical gap
\cite{gamma}.
\par
\begin{figure}
\includegraphics[width=6cm]{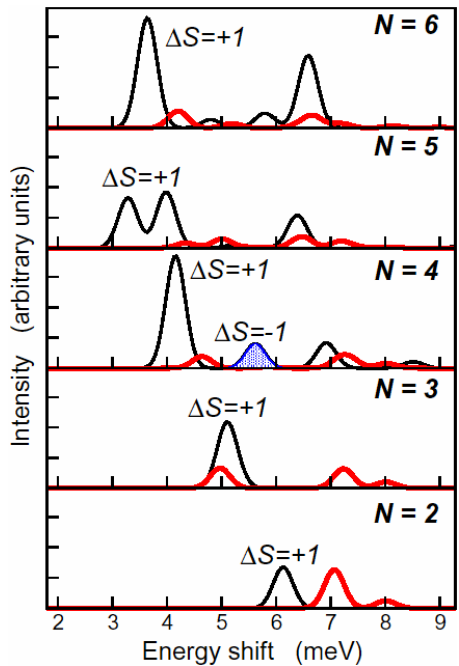}
\caption{\label{fig2} Evolution of the theoretical inelastic light
scattering spectra as a function of electron occupation number $N$
within a configuration-interaction approach. Red (black) curves
represent charge (spin) excitations. The blue peak is the TS
monopole mode.}
\end{figure}

Figure \ref{fig1}(b) displays the calculated spectra for $N=4$ and
$\hbar\omega _0 = 4$ meV. The latter value was determined by
fitting the peak energy position in the
experimental spectra shown in Fig.~\ref{fig1}(a). An independent
check for this value of $\omega_0$ and $N$ comes from the
empirical relation given in Ref.~\onlinecite{Reimann} [Eq.~(11)]
linking $N$, $\omega_0$, and the electron density, $n_e$, which
gives $n_e=1.2\times 10^{11}$ cm$^{-2}$, in good agreement with
the experimental value. It can be seen that among all calculated
excitations with $\Delta M=0$, only a few of them turn out to have
significant intensities, generating discrete spectrum lines (with
a phenomenological broadening
%standard deviation $\sigma  = $0.18 meV
chosen to reproduce the measured TS linewidth) in very good agreement
with the experimental ones. It can also be noted that more than
one excitation gives a significant contribution to the spectra at
energies above the TS mode. This is consistent with the observed
larger linewidths for the higher-energy excitation pairs. Figure
\ref{fig2}(a) reports the evolution of the calculated spectra as a
function of $N$. As expected, the TS ($\Delta S = -1$) mode is 
peculiar to $N=4$ and it is not
observed at any other explored electron occupation configurations.

\par
The excitations of Fig.~\ref{fig2} show a
redshift of the lowest-energy features in both channels as $N$
(and $n_e$) increases \cite{brocke} due to screening effect.
Because of the exchange energy gain of excited states, the spin
channel energy is systematically lower than the charge excitation
energy. This large sensitivity of the light scattering spectra on
particle occupation is at the origin of the difference between the
observed linewidths of our inter-shell excitations. Comparing the
evolution of peak energies shown in Fig. 2 with measured
linewidths we conclude that a distribution of electron occupation
between $4$ and $6$ characterizes our QD arrays. It also indicates that
the light scattering method allows to probe excitation spectra of
few-electron QDs with single-electron
accuracy despite the relatively large number of QDs illuminated.
Consistent with the assignment that links the $\Delta S=-1$ mode
to those selected QDs that have four electrons, is the observed
sharp linewidth of 0.4 meV which is much lower than the linewidths
of the other spin and charge transitions.
\par
\begin{figure}
\includegraphics[width=7cm]{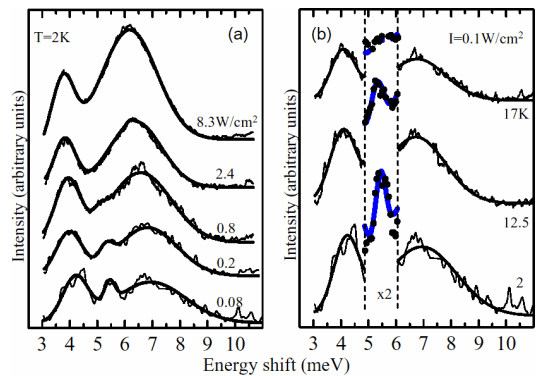}
\caption{\label{fig3} (a) Evolution of spin excitations as a
function of incident laser intensity.
The resonant laser energy is 1567 meV.
(b) Temperature dependence of spin transitions.
The laser intensity I is 0.1 W/cm$^2$.
Spectra are presented after conventional subtraction of
background due to laser light.}
\end{figure}

The evolution of the spin transitions at different incident laser
intensities shown in Fig.~\ref{fig3}(a) confirms that the QDs are in
the few-electron regime. As the intensity increases we expect
additional electrons to be photo-generated. Consistent with Fig.~2, 
we found that the peaks display a
red shift and that the TS transition disappears at around I = 1
W/cm$^2$, suggesting that at this intensity all the QDs have more
than 4 electrons and therefore the number of those photo-generated
is at least one. In addition, contrary to the other inter-shell
modes the intensity of the TS  spin excitation
decreases significantly as temperature increases
[Fig.~\ref{fig3}(b)] with an estimated activation gap of $0.7 \pm
0.3$ meV. At such low energy a possible thermally populated excited
level is the singlet state without any change in orbital occupation.
This energy thus provides an estimate of the low lying intra-shell
singlet-triplet transition of the four-electron QDs and it compares
well with CI estimate of 0.8 meV.
\par
\par
\begin{figure}
\includegraphics[width=7cm]{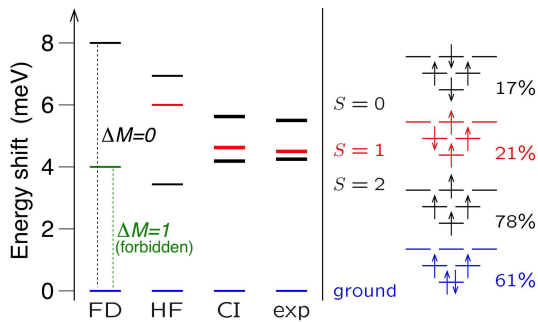}
\caption{\label{fig4} Comparison between measured (right column)
and calculated excitation energies of charge (red) and spin
(black) channels, identified by arrows in Fig. 1:
From left to right, non-interacting Fock-Darwin model (FD), self-consistent
Hartree-Fock (HF), full configuration interaction (CI).
Side diagrams show the most-weighted configurations in the CI linear
expansion of correlated wave functions, with the corresponding weight
percentage. HF calculations refer to such configurations.}
\end{figure}

A specific feature of the low-$N$ regime studied here is that states
are represented by superpositions of many different SDs to
incorporate both radial and angular spatial correlation
\cite{Steinbach}. The side diagrams of Fig.~\ref{fig4} represent the
weighted SDs in the CI expansion of the $N=4$ ground and excited
states involved in the three transitions indicated by arrows in
Fig.~1(b). We depicted the states corresponding to the maximum
allowed $S_z$, while in the actual calculation we only considered
the degenerate states with $S_z=0$ \cite{method}.
\par
\par
Figure 4 also shows ground- and excited-state energies
calculated with different approximations 
that provide evidence for  correlation effects in the
excitation spectra. In the FD picture the energy
difference between consecutive levels is given by $\hbar \omega_0 $
= 4 meV. In the HF approach, spin-orbitals are computed
self-consistently \cite{Rontani99}. The energy difference between
the three spin configurations is given by bookkeeping the exchange
energy $K_{ab}$ gained each time two electron spins, occupying any
orbitals $a$ and $b$, are parallel to each other. This gain is
accounted for by the Coulomb exchange integral between orbitals $a$
and $b$ described by FD wave functions. This approach neglects
spatial correlation among electrons.
\par
Correlation effects are included in the CI approach, 
leading to the theoretical spectra in Figs.~1 and 2 and to
the quantitative agreement with experiments shown in Fig.~4.
The comparison between HF and CI  (Fig.~4) suggests that
correlation affects the 
relative splittings between excited states, 
even reversing their relative amplitudes: The $S=1$ state
is nearer to $S=0$ than to $S=2$ in HF, while the opposite is
true in CI, in agreement with the experiment. As suggested by the decreasing
contribution of the most weighted SD configurations indicated on the
right in Fig.~4, correlation effects are small for the ground
and the $S=2$ excited state, but become increasingly
important for excited states with smaller $S$. As $S$ decreases,
exchange interaction is less effective in keeping electrons far
apart and excited states become more correlated. Note that
the relative amplitudes of the calculated HF and CI gaps are
quite insensitive to the specific value of $\hbar\omega_0$ and we found that 
the measured splittings among the spin modes can only be reproduced by 
CI calculations, no matter the value of $\hbar\omega_0$.
\par
In conclusion, we reported inelastic light scattering
measurements of spin transitions in nanofabricated quantum dots.
The characteristic excitations of the triplet
configuration with four electrons have been identified and
theoretically evaluated. We have shown that light scattering 
methods offer a wealth of information on
the physics of spin states in QDs with few electrons.
 \par
We acknowledge support from the Italian Ministry of Foreign
Affairs, Italian Ministry of Research (FIRB-RBAU01ZEML and
COFIN-2003020984), CINECA-INFM Supercomputing Project 2005,
European Community's Human Potential Programme
(HPRN-CT-2002-00291), National Science Foundation (DMR-03-52738),
Department of Energy (DE-AIO2-04ER46133) and a research grant of
the W. M. Keck Foundation. We are grateful to SENTECH-Berlin for
allowing us to use the ICP-RIE machine. We thank F. Beil, J.P.
Kotthaus, and F. Troiani for discussions.


\begin{thebibliography}{99}
\bibitem{gamma}
{At difference with Ref.~\onlinecite{Steinebach}, 
we neglected Coulomb interaction between
electrons and heavy holes and we considered a
finite-width quantum well for both electrons
and holes. Spectra were computed under resonance
conditions, with laser energy larger than optical gap by an amount
of 8 meV, and $q=1\times 10^4$ cm$^{-1}$.}
\end{thebibliography}
\end{document}